\documentclass[conference]{IEEEtran}
\IEEEoverridecommandlockouts

\usepackage{cite}
\usepackage{amsmath,amssymb,amsfonts}
\usepackage{graphicx}
\usepackage{textcomp}
\usepackage{xcolor}
\usepackage{algorithm}
\usepackage{algpseudocode}
\usepackage{comment}
\usepackage{url}
\usepackage{booktabs}
\usepackage{multirow}
\usepackage{hyperref}
\usepackage{orcidlink}
\usepackage{mathtools}
\usepackage{lipsum}  
\DeclarePairedDelimiter{\ceil}{\lceil}{\rceil}

\def\BibTeX{{\rm B\kern-.05em{\sc i\kern-.025em b}\kern-.08em
    T\kern-.1667em\lower.7ex\hbox{E}\kern-.125emX}}

\makeatletter
\newcommand{\linebreakand}{
  \end{@IEEEauthorhalign}
  \hfill\mbox{}\par
  \mbox{}\hfill\begin{@IEEEauthorhalign}
}
\makeatother

\makeatletter
\newenvironment{breakablealgorithm}
  {
   \begin{flushleft}
     \refstepcounter{algorithm}
     \hrule height.8pt depth0pt \kern2pt
     \renewcommand{\caption}[2][\relax]{
       {\raggedright\textbf{\fname@algorithm~\thealgorithm} ##2\par}%
       \ifx\relax##1\relax 
         \addcontentsline{loa}{algorithm}{\protect\numberline{\thealgorithm}##2}%
       \else 
         \addcontentsline{loa}{algorithm}{\protect\numberline{\thealgorithm}##1}%
       \fi
       \kern2pt\hrule\kern2pt
     }
  }{
     \kern2pt\hrule\relax
   \end{flushleft}
  }
\makeatother

\begin{document}

\title{Optimised Quantum Embedding: A Universal Minor-Embedding Framework for Large Complete Bipartite Graph}

\author{\IEEEauthorblockN{Salvatore Sinno \orcidlink{0009-0002-9177-5161}\IEEEauthorrefmark{1}\\
Thomas Gro{\ss} \orcidlink{0000-0002-7766-2454
 },\IEEEauthorrefmark{1} Nicholas Chancellor \orcidlink{0000-0002-1293-0761}\IEEEauthorrefmark{1}\\
 Bhavika Bhalgamiya \orcidlink{0009-0001-8586-4531}, \IEEEauthorrefmark{2} and   Arati Sahoo
 \IEEEauthorrefmark{2}}\\

\IEEEauthorblockA{\IEEEauthorrefmark{1}School of Computing, Newcastle University, Newcastle upon Tyne, United Kingdom\\
Email: \{S.Sinno2\}@newcastle.ac.uk}

\IEEEauthorblockA{\IEEEauthorrefmark{2}NextGen Computing Research Group, Unisys, Blue Bell, Pennsylvania, USA\\}
}
\maketitle

\begin{abstract}
Minor embedding is essential for mapping large-scale combinatorial problems onto quantum annealers, particularly in quantum machine learning and optimization. This work presents an optimized, universal minor-embedding framework that efficiently accommodates complete bipartite graphs onto the hardware topology of quantum annealers. By leveraging the inherent topographical periodicity of the physical quantum adiabatic annealer processor, our method systematically reduces qubit chain lengths, resulting in enhanced stability, computational efficiency, and scalability of quantum annealing.

We benchmark our embedding framework against Minorminer, the default heuristic embedding algorithm, for the Pegasus topology, demonstrating that our approach significantly improves embedding quality. Our empirical results show a 99.98\%  reduction in embedding time for a 120×120 complete bipartite graph. Additionally, our method eliminates long qubit chains, which primarily cause decoherence and computational errors in quantum annealing. 
These findings advance the scalability of quantum embeddings, particularly for quantum generative models, anomaly detection, and large-scale optimization tasks. Our results establish a foundation for integrating efficient quantum-classical hybrid solutions, paving the way for practical applications in quantum-enhanced machine learning and optimization.
\end{abstract}


\section{Introduction}

Adiabatic quantum computing systems based on superconducting qubits impose strict physical constraints on hardware connectivity. Specifically, each qubit is restricted in terms of both its degree, limiting the number of direct couplers and the physical length of couplers, restricting connections to nearby qubits only. Although the physical arrangement of qubits can be adapted to some degree, practical quantum hardware graphs are fundamentally bounded-degree, edge-length-constrained geometric layouts.

Embedding logical problems, such as the Ising Hamiltonian, onto these hardware architectures is thus constrained by these physical limitations. If the logical graph is not directly representable within the hardware's connectivity constraints, a technique known as minor embedding must be applied. Minor embedding involves mapping logical qubits onto interconnected physical qubit chains, ensuring logical connectivity by creating subtrees that effectively behave as single logical qubits.

The impact of minor embedding extends beyond just execution time—it also affects the quality of the solution. In many cases, to achieve the required connectivity between qubits, the embedding process forces the creation of long chains of physical qubits to represent a single logical qubit. These chains, which correspond to higher-order neighbourhoods in the quantum hardware graph, introduce structural vulnerabilities. When chains exceed a critical length threshold (typically more than six qubits long), they become increasingly prone to breakage due to thermal noise, fabrication imperfections, and quantum decoherence. This leads to unreliable computational results, as broken chains disrupt the logical qubit representation and degrade the overall performance of the quantum algorithm.

Therefore, optimizing the minor embedding process is essential for ensuring efficient use of AQC resources, reducing computation overhead, and improving the accuracy and stability of the final solutions obtained from quantum annealing.

\textbf{Our Contribution:} This paper introduces an optimized minor embedding framework designed for mapping large complete bipartite graphs (e.g., Restricted Boltzmann Machines  (RBMs)) onto adiabatic quantum processors. Our main contributions are:

\begin{enumerate}
\item A universal minor-embedding framework tailored for efficiently embedding large complete bipartite graphs onto quantum annealers, notably leveraging the Pegasus topology.
\item A significant reduction in qubit chain lengths, thus enhancing embedding stability and computational performance compared to established embedding algorithms, including Minorminer.
\item Empirical evaluations demonstrate that our embedding approach enables faster embedding (at least two orders of magnitude faster than other metaheuristic approaches)
\end{enumerate}

\section{Related Work}
A scalable superconducting architecture for adiabatic quantum computation (AQC) was initially proposed by Kaminsky \emph{et al.} \cite{Kaminsky2004}, based on the transverse-field Ising Hamiltonian. D-Wave Systems Inc. subsequently developed superconducting quantum processors using flux qubits connected via tunable coupling devices \cite{Dwave2009}. Despite these hardware advancements, embedding complex optimization problems, including complete bipartite graphs, onto such quantum processors remains a challenging task in AQC.

Early embedding techniques, such as the TRIAD method \cite{choi2008minor, choi2011minor}, were designed to efficiently embed complete graphs ($K_n$) by mapping each vertex to chains of qubits arranged in regular patterns. Although these methods minimize coupler overhead for fully connected topologies, they often provide suboptimal embeddings for bipartite structures (i.e., complete bipartite graphs). In particular, the chain-based minor embeddings typically grow in length for bipartite graphs, increasing vulnerability to decoherence and chain breakage \cite{choi2008minor, choi2011minor}.

Recent minor-embedding strategies have focused on optimizing hardware efficiency specifically for bipartite graphs, leveraging the structured layouts found in modern quantum architectures \cite{serra2021template,date2021efficiently}. Serra \emph{et al.} \cite{serra2021template} demonstrated a template-based approach that significantly reduces embedding overhead, and Goodrich \emph{et al.} \cite{goodrich2018optimizing} alongside Zbinden \emph{et al.} \cite{zbinden2021embedding} presented systematic heuristics that achieve practical reductions in chain length for bipartite embeddings, particularly on Pegasus topologies. These advances highlight the importance of architecture-specific and graph-specific optimization strategies to ensure efficient utilization of quantum resources.

An important practical application of complete bipartite graphs arises in Restricted Boltzmann Machines (RBMs), which have received considerable attention in Quantum Machine Learning (QML) due to their ability to model complex probability distributions. Quantum Annealing approaches for RBM training have shown promise in generative modelling and anomaly detection \cite{benedetti2023anomaly,adachi2015application,benedetti2016estimation}. In these scenarios, Quantum RBMs (QRBMs) leverage the Ising Hamiltonian to implement the bipartite structure directly on quantum hardware, reducing simulation overhead and potentially enhancing performance \cite{dixit2021training,klymko2014adiabatic}. Consequently, the choice of quantum processor topology, such as Chimera or Pegasus, critically impacts the scalability and accuracy of QRBMs by affecting qubit connectivity and embedding complexity \cite{date2021efficiently,klymko2014adiabatic}.

Benedetti \emph{et al.} \cite{benedetti2023anomaly} underscored the necessity of tailored embedding approaches for specific quantum architectures when deploying RBMs. Building on these insights, our proposed universal embedding framework seeks to address the limitations in existing minor embeddings for complete bipartite graphs. By systematically exploiting the enhanced connectivity in modern quantum processors, our approach ensures both scalable and robust embeddings of bipartite structures, thereby offering a practical pathway for quantum-enhanced generative modelling and anomaly detection tasks.

\section{Theoretical Analysis}
\subsection{Optimal Embedding of Complete Bipartite Graphs on Pegasus}
This section presents a theoretical analysis of the optimal embedding for complete bipartite graphs on quantum annealing architectures.
A complete bipartite graph is a unique graph divided into two distinct node groups (or sets). Each node in one set is connected to every node in the other set, but there are no connections between nodes within the same set. A Restricted Boltzmann Machine (RBM) is an example of a complete bipartite graph, as it has two sets of units—visible and hidden—with each visible node fully connected to all hidden nodes and no internal connections within the same layer.

An adiabatic quantum architecture can represent an undirected graph $U$ with weighted vertices and edges. Each vertex $i \in V(U)$ corresponds to a qubit, and each edge $ij \in E(U)$ corresponds to a coupler between qubits $i$ and $j$. Throughout the discussion, we will use \emph{qubit} and \emph{vertex}, as well as \emph{coupler} and \emph{edge}, interchangeably when there is no ambiguity.
Each qubit $i$ has two associated weights: $h_i$ (the qubit bias) and $\Delta_i$ (the tunnelling amplitude). Similarly, each coupler $ij$ has an associated weight $J_{ij}$ (the coupler strength). These weights are time-dependent functions, typically expressed as $h_i(t)$, $\Delta_i(t)$, and $J_{ij}(t)$, to accommodate dynamic control during computation. Setting $J_{ij} = 1$ locks the two qubits together (antiferromagnetic coupling): this allows the formation of chains and extends the connectivity of physical qubits to qubits beyond the rank-one neighbourhood (first-order neighbourhood) consisting of all qubits that are directly connected to the given vertex by a single edge. In the case of an adiabatic quantum computing architecture, rank-one neighbourhoods define which qubits are directly coupled in a given hardware graph.

\begin{figure}[hbt!]
\caption[]{Minor-embedding of G(left) in a square lattice U.}
\label{fig:GandU}
\begin{center}
    \includegraphics[width=0.4\textwidth]{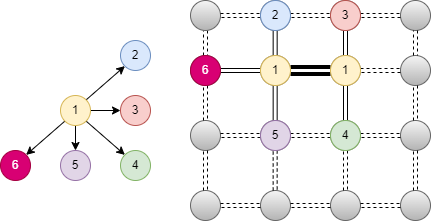}
\end{center}
\end{figure}

Fig \ref{fig:GandU} we show how a $G^{emb}$ is created as a minor embedding of a graph $G$ on the lattice $U$ of the physical topology. Each vertex of the logical graph $G$ is mapped against a subtree and vertices of the graph $U$. $G$ is called a graph minor of $U$ \cite{choi2010minor}. 

The optimal embedding of complete bipartite graphs, $\epsilon: K_{a,b} \rightarrow P$, consists of a parallel path of horizontal qubits and a parallel path of vertical qubits.

To identify the theoretical optimal embedding, we start with the actual Pegasus architecture and then generalise the approach.
In the Pegasus architecture, each qubit has a nominal length of 12 and a degree of 15. The relation $D=12M(M-1)$ gives the total qubits in Pegasus, where $M$ is the number of vertical and horizontal cells. $ P_M$ therefore identifies Pegasus architectures with $M$ vertical and horizontal cells. The theoretical maximum embedding of a complete graph (\textit{clique}) is the largest clique embeddable is $K_{12M-10}$ \cite{boothby2020nextgenerationtopologydwavequantum}.
The theoretical maximum for complete bipartite graph ($K_{m,n}$) is given by \cite{boothby2020nextgenerationtopologydwavequantum}:
\begin{equation}
\label{eq:optima_embeddign_size}
  max(K_{m,m})=K_{12M-20,12M-20}
\end{equation}
This is because bipartite graphs require fewer edges per node, thus allowing for larger embedding than complete cliques. In practical application, embedding overhead (chain of qubits representation logical nodes) reduces the practical maximum size.
So for $P_{16}$, the maximum RBM theoretically embedded given by Eq. \ref{eq:optima_embeddign_size} is 172$\times$172.

\subsection{Our Embedding Algorithm}
This section presents a generalized definition of quantum adiabatic processor topologies based on a generic family of structured graphs, denoted as $G(M)$, where $M$ determines the number of qubits in the architecture. In other words, a linear function provides the maximum number of qubits $D$ for a given $M$:

\begin{equation}
D = \alpha M(M-1)
\end{equation}

where $\alpha$ is a parameter defining the number of qubits directly connected to each qubit (in other words, $\alpha$ is the degree of the vertex).

Qubits are arranged in vertical and horizontal orientations in the considered generic topology, similar to established topologies like Chimera and Pegasus. These qubits are grouped into cells of fully connected vertices ($K_n$), forming a complete bipartite graph $K_{a,b}$.

This topology can be visualized as a grid, where each qubit has a unique identifier based on its orientation (horizontal or vertical) and the tile to which its cell is allocated. Based on this, we define a coordinate system for qubits in $G$. A qubit $(u, w, k, z)$ in $G$ is described by four parameters: $u$ represents the orientation, where $u=0$ for vertical and $u=1$ for horizontal qubits. The perpendicular tile offset $w$ indicates the index of the qubit's tile in the direction perpendicular to $u$, meaning $w$ is a horizontal (column) index if $u=0$ and a vertical (row) index if $u=1$. The qubit offset $k$ denotes the position of a qubit within a tile. The parallel tile offset $z$ specifies the index of the qubit's tile in the direction parallel to $u$, such that $z$ is a vertical (row) index when $u=0$ and a horizontal (column) index when $u=1$.

The integer labelling for $G$ is defined as:

\begin{equation}
\label{eq:coordinates_id_qubit}
(u, w, k, z) \mapsto z+(M-1)(k+\alpha (w+M u))
\end{equation}

\begin{figure}[hbt!]
\caption[]{Connections}
\label{fig:qubit_connections}
\begin{center}
     \includegraphics[width=0.3\textwidth, height=0.3\textwidth]{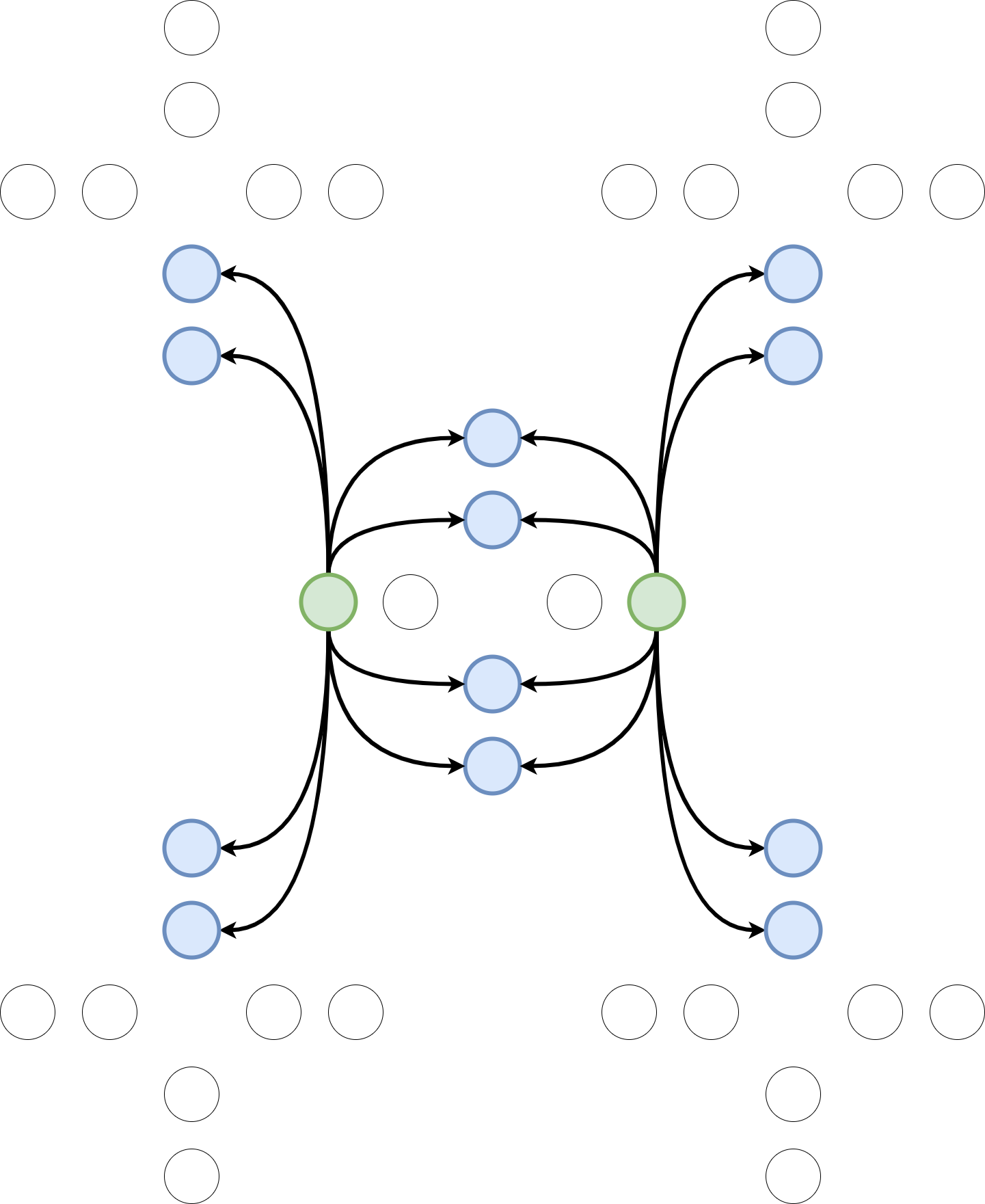}
\end{center}
\end{figure}

Each left horizontal qubit (\( u = 0 \)) is connected to all vertical qubits within its own cell, as well as to the bottom and top qubits in the adjacent cell to the right. Similarly, each right horizontal qubit (\( u = 1 \)) is connected to all vertical qubits within its cell and the bottom and top vertical qubits in the neighbouring cell to the right (see Figure \ref{fig:qubit_connections}). Additionally, the external connections \( (u, w, k, z) \sim (u, w, k, z+1) \) ensure that each qubit maintains connectivity across the lattice structure.

Based on this, to embed a complete bipartite graph $K_{V,H}$, equivalent to an RBM with $V$ visible nodes and $H$ hidden nodes, we start from a completed bipartite subgraph that can be embedded using some $K_{m,n}$ bipartite graph that can be completed embedded in $G(M)$ using $K_{a,b}$.
If $V \le min(m,n)M$, then the number of vertical $K_{m,n}$ cells we need is given by: 

\begin{equation}
H_v=\ceil[\bigg]{\frac{V}{min(m,n)}}
\end{equation}
In this case, each visible node $k$ belonging to a cell $(w, k, z)$ is mapped to a physical horizontal qubit given by the relation: 
\begin{equation}
\label{eq:qubit_V_id}
Vqubit_{id} = startV_{id}+\alpha \times k+z \times (\alpha \times w)
\end{equation}

$startV_{id}$ is given by Eq. \ref{eq:coordinates_id_qubit} replacing $(u,z,k,w)$ with $(0,0,0,1)$.
Similarly, for the hidden qubits, mapping the hidden qubits to a vertical qubits, by the relation:

\begin{equation}
\label{eq:qubit_H_id}
Hqubit_{id} = startH_{id}+\alpha \times k+z \times (\alpha \times (w+M))
\end{equation}
$startH_{id}$ is given by Eq. \ref{eq:coordinates_id_qubit} replacing $(u,z,k,w)$ with $(1,0,0,1)$.

If $V > min(m,n)M$, for the first $min(m,n)M$ visible(hidden) units, we use the equations \ref{eq:qubit_V_id} and \ref{eq:qubit_H_id}, and we embed the other visible (hidden node) using additional cells given by

\begin{equation}
H_{v,add}= V\mod min(m,n)
\end{equation}
In this case \ref{eq:qubit_V_id} and \ref{eq:qubit_H_id} still applies, however we have:
$startV_{id}$ is given by Eq. \ref{eq:coordinates_id_qubit} replacing $(u,z,k,w)$ with $(0,0,1,1)$ and $startH_{id}$ is given by Eq. \ref{eq:coordinates_id_qubit} replacing $(u,z,k,w)$ with $(1,1,0,1)$.
This algorithm only requires that the nominal length of the QAC processor is $alpha \le m+n+1$ to embed this structure in a physical architecture, the physical identifies of each qubit can be identified by \ref{eq:coordinates_id_qubit} and the $K_{m,n}$ complete bipartite graph can embedded as shown in Fig \ref{fig:qubit_connections}.
Our embedding requires $O(mn/d)$ physical qubits since the edges only exist between visible and hidden, and no intra-layer coupling reduces the number of virtual qubits per logical qubits.
The maximum RBM (or complete bipartite graph) that can be embedded is therefore $max(Hidden) = (M-1)n$  $max(H) \le \alpha M$ and $max(Visible) = (3M-2)m$ with $max(Visible) \le \alpha M$. The maximum complete bipartite graph that can be minor embedded on the Pegasus architecture using this algorithm is $K_{172,120}$.

\section{Empirical Embedding for RBM}
\begin{figure}[hbt!]
\caption[Mapping RBM to D-Wave]{Mapping RBM to D-Wave}
\label{fig:mapping_restricted_qpu_1}
\begin{center}
    \includegraphics[width=0.5\textwidth, height=15cm]{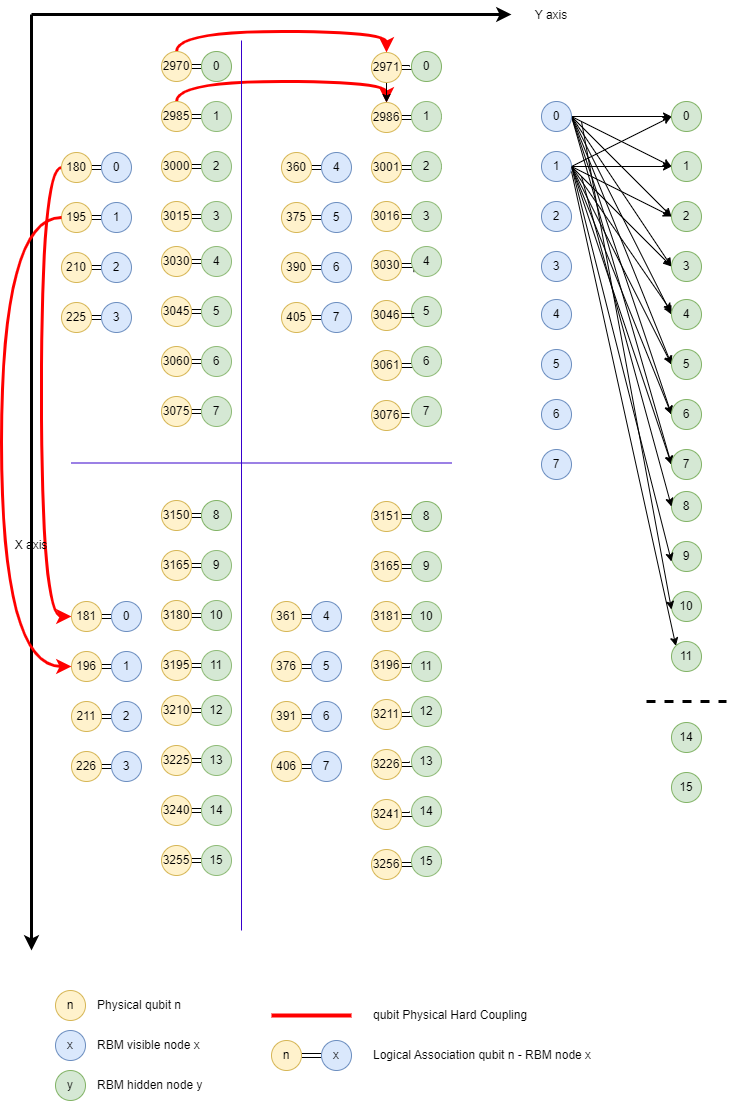}
\end{center}
\end{figure}

For the Pegasus architecture with $l=12$, we select $m$ and $n$ as 4 and 8
Following our algorithm (Figure \ref{fig:mapping_restricted_qpu_1}), each physical qubit is assigned a corresponding logical identifier in embedding a Restricted Boltzmann Machine (RBM) onto a quantum processing unit (QPU). In this mapping, visible nodes are represented in blue, hidden nodes in green, and physical qubits (yellow) are uniquely identified by an ID. 
For illustrative purposes, in this example, we embed an 8$\times$16 RBM. To establish connectivity between visible and hidden nodes, specific qubits are coupled to ensure continuity in the mapping. For example, qubit 180 is mapped to visible node 0. Within the QPU architecture, visible node 0 connects to a set of physical qubits with identifiers 2970, 2985, 3000, 3015, 3030, 3045, 3060, and 3075. These qubits are logically assigned to hidden nodes 0 through 7, ensuring that visible nodes 1, 2, and 3 are linked to the same set of hidden nodes.

Additional couplings are introduced to extend this connectivity further. Visible nodes 4, 5, 6, and 7 are connected to the same hidden nodes (0–7) using a structured coupling mechanism. For instance, qubit 2970 is coupled with qubit 2971, qubit 2985 with qubit 2986, and so on, continuing in this pattern until qubit 3075 is coupled with qubit 3076. This arrangement maintains logical consistency, ensuring visible units 0–7 are correctly mapped to hidden units 0–7.

Vertical couplings are introduced to expand the number of hidden units. Qubit 180 is coupled with qubit 181, allowing visible node 0 to connect to an additional set of hidden nodes (8–15), which are mapped to physical qubits ranging from 3150 to 3255 in increments of 15.

Similarly, to extend the connections of visible nodes 4, 5, 6, and 7 to hidden nodes 8–15, additional couplings are established: qubit 3150 is coupled with qubit 3151, qubit 3165 with qubit 3166, and qubit 3255 with qubit 3256.

This structured coupling mechanism provides a scalable and systematic approach for mapping visible and hidden nodes in the RBM. This method supports implementing RBMs on quantum annealers by maintaining logical connectivity and ensuring efficient qubit utilisation.

\begin{figure}[hbt!]
\caption[8$\times$8 RBM Embedded on Pegasus]{8$\times$8 RBM Embedded on Pegasus}
\label{fig:8x8_RBM_Embedded_Pegasu}
\begin{center}
    \includegraphics[width=0.5\textwidth, height=0.5\textwidth]{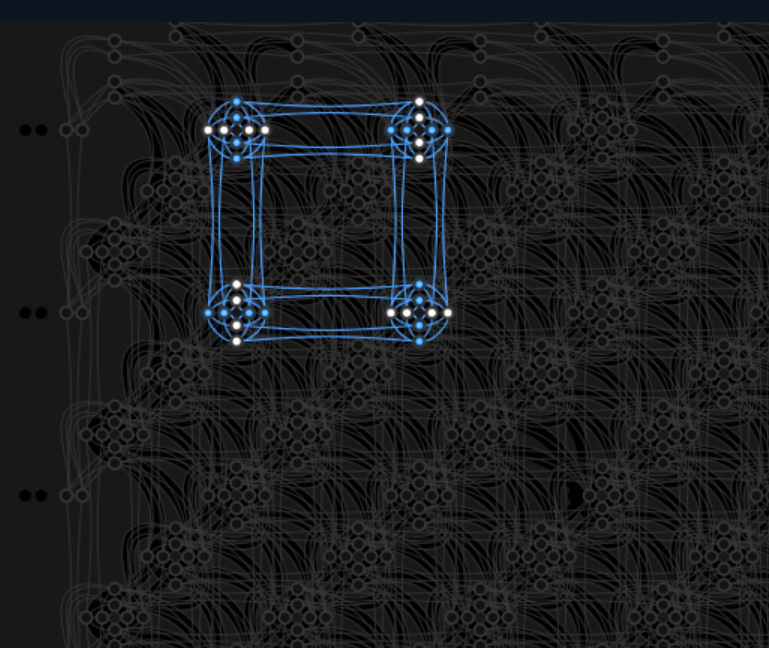}
\end{center}
\end{figure}

\begin{figure}[hbt!]
\caption[8$\times$12 RBM Embedded on Pegasus]{8$\times$12 RBM Embedded on Pegasus}
\label{fig:8x12_RBM_Embedded_Pegasu}
\begin{center}
    \includegraphics[width=0.5\textwidth, height=0.5\textwidth]{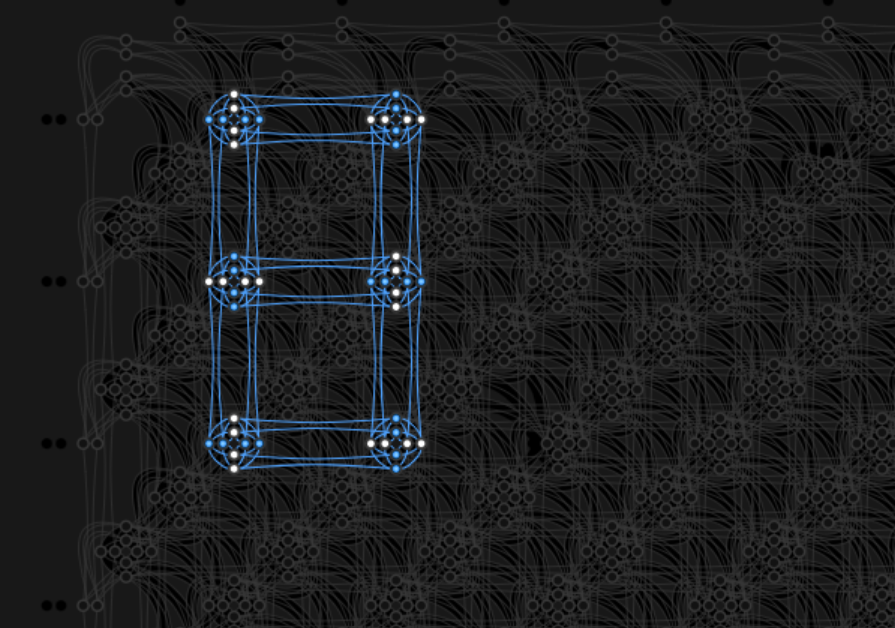}
\end{center}
\end{figure}

Figures \ref{fig:8x8_RBM_Embedded_Pegasu} and \ref{fig:8x12_RBM_Embedded_Pegasu} illustrate this embedding process within the Pegasus architecture for an 8$\times$8 and 8$\times$12 Restricted Boltzmann Machine (RBM), respectively.

In the case of an 8$\times$8 RBM, the embedding requires a 2$\times$2 grid of unit cells. Each horizontal row cell contributes four units to the hidden nodes, while each vertical column cell adds four units to the visible nodes.

For an 8$\times$12 RBM, the required structure expands to two vertical columns and three horizontal rows to accommodate the additional hidden and visible nodes.

\begin{figure}[hbt!]
\caption[44$\times$44 RBM Embedded on Pegasus]{44$\times$44 RBM Embedded on Pegasus}
\label{fig:44x44_RBM_Embedded_Pegasu}
\begin{center}
    \includegraphics[width=0.5\textwidth, height=0.5\textwidth]{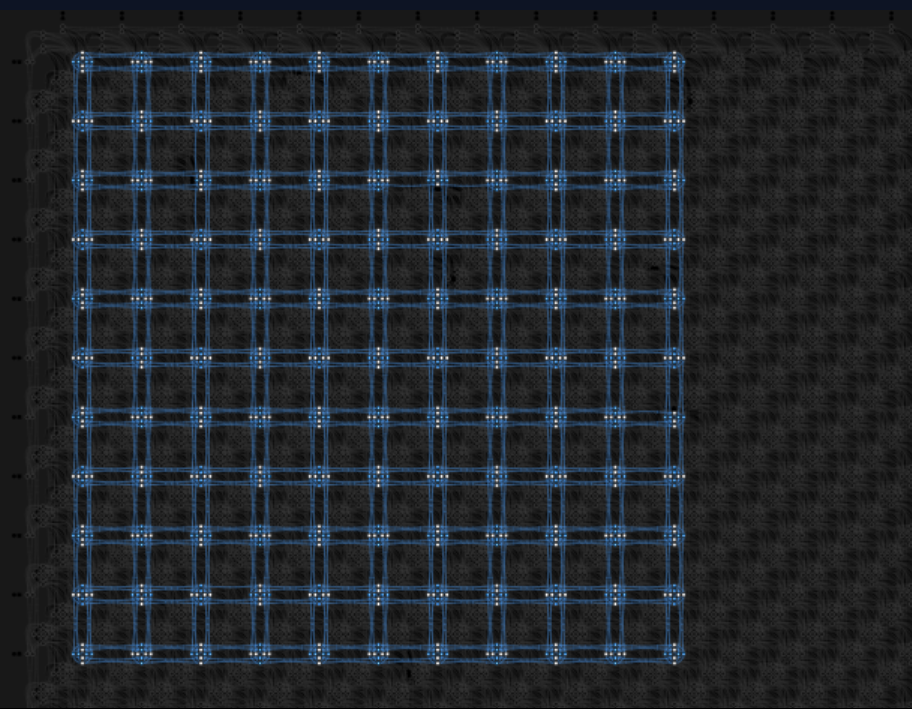}
\end{center}
\end{figure}

Figure \ref{fig:44x44_RBM_Embedded_Pegasu} illustrates the embedding of a \( 44 \times 44 \) Restricted Boltzmann Machine (RBM) within the Pegasus architecture. Given the structured approach used for smaller RBMs, such as the \( 8 \times 8 \) and \( 8 \times 12 \) cases, the embedding for a \( 44 \times 44 \) RBM follows a similar pattern but on a significantly larger scale.

As a result, the \( 44 \times 44 \) RBM requires an 11 $\times$ 11 grid.

\begin{figure}[hbt!]
\caption[60$\times$60 RBM Embedded on Pegasus]{60$\times$60 RBM Embedded on Pegasus}
\label{fig:60x60_RBM_Embedded_Pegasu}
\begin{center}
    \includegraphics[width=0.5\textwidth, height=0.5\textwidth]{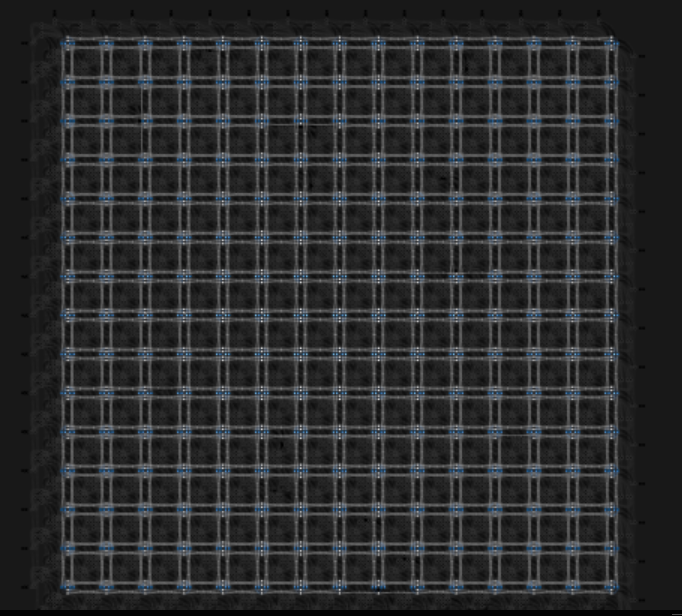}
\end{center}
\end{figure}

Figure \ref{fig:60x60_RBM_Embedded_Pegasu} illustrates the embedding of a \( 60 \times 60 \) Restricted Boltzmann Machine (RBM) within the Pegasus architecture. Following the structured approach used for smaller RBMs, such as the \( 8 \times 8 \), \( 8 \times 12 \), and \( 44 \times 44 \) cases, the embedding for a \( 60 \times 60 \) RBM scales accordingly while preserving connectivity and logical qubit assignments.

As a result, the \( 60 \times 60 \) RBM requires an 15 $\times$ 15 grid of Pegasus unit cells to ensure full connectivity between visible and hidden nodes. This embedding structure efficiently distributes the logical-to-physical mapping across the available qubits while preserving the connectivity required for successful quantum annealing.


\begin{figure}[t]
\caption{Embedding 120$\times$120 RBM on Pegasus}
\includegraphics[width=0.5\textwidth]{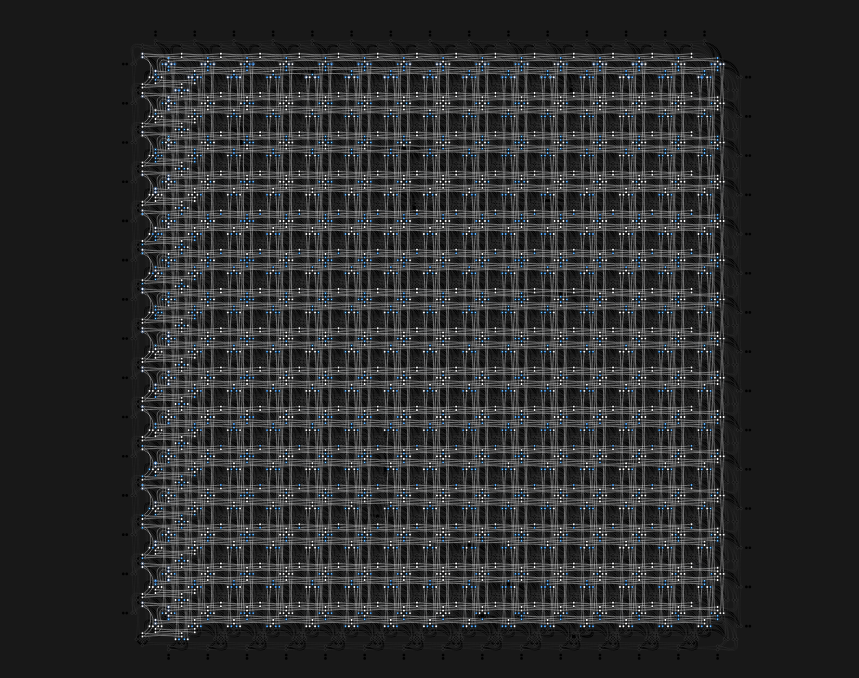}
\label{fig:Embedding_120_120}
\end{figure}

\begin{figure}[t]
\caption{Deail of Embedding 120$\times$120 RBM on Pegasus}
\includegraphics[width=0.5\textwidth]{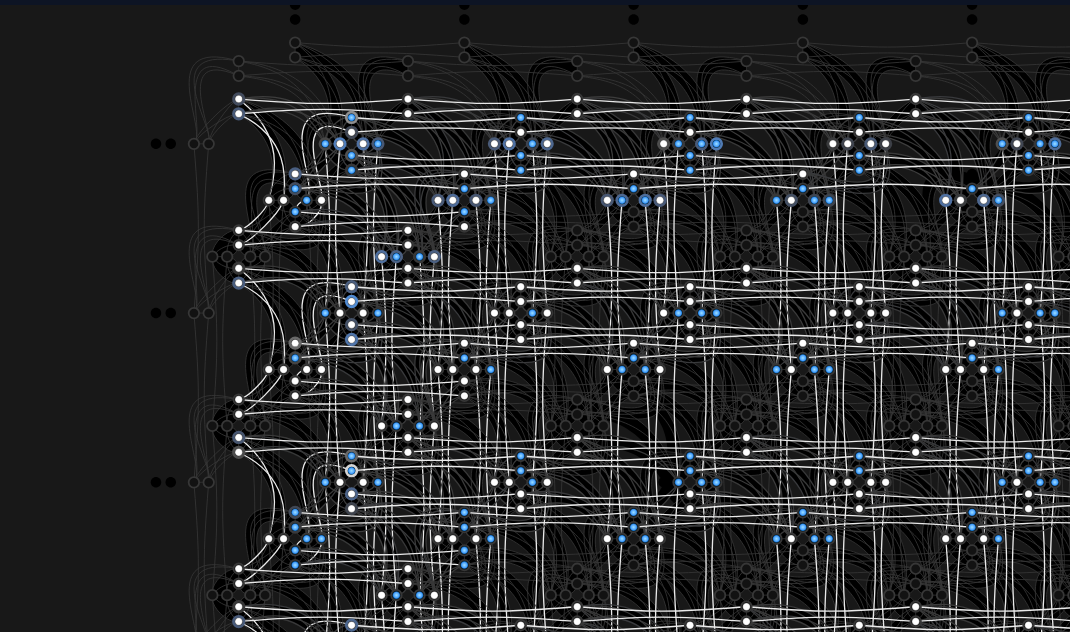}
\label{fig:Detail_120_120_embedding}
\end{figure}

\section{Experimental Results}
This section compares our embedding with the default heuristic embedding of Mminorminer.

\subsection{Methodology}
To compare our algorithm with the default heuristic embedding, we have considered RBM to embed complete bipartite graphs $K_{n,n}$ with $n=40$ to $n=120$ in increments of 20.
For a fixed number of trials $T$ equal to 100 for each configuration, we define two metrics:
\begin{itemize}
    \item \textit{average time to embed (s)}: This represents the average time required to produce the embedding
    \item \textit{the average number of chains with length $\geq$ 6}: this represent the chains longer than six. Chain of length six or above are subjected to breaking, and therefore,e embedding with a chain length of 6 or more produces unreliable results
\end{itemize}

The experiments were conducted using a hybrid setup. Our embedding algorithm was used to run a Dell R705X server with 128 logical cores and 503 GiB of RAM. The server ran Ubuntu 22.04 and Python 3.10.9. Minorminer 0.2.16 was used to compare the results with the default D-Wave embedding algorithm.

Our embedding algorithm (Algorithm \ref{alg:embedding_graph}) allows for shorter qubit chains, reducing the risk of chain breakage and improving the accuracy of the annealing process and outperforms the heuristic Minorminer algorithm as shown in Table \ref{tab:embedding_comparison}. In contrast, our algorithm efficiently identifies multiple RBM configurations in milliseconds. Once the embedding is generated, the graph is passed to the annealer using the 
$J$ and $h$ matrices, ensuring seamless integration into the quantum computation process.

\begin{figure}[h!]
\caption{Comparison of Embedding}
\includegraphics[width=\columnwidth]{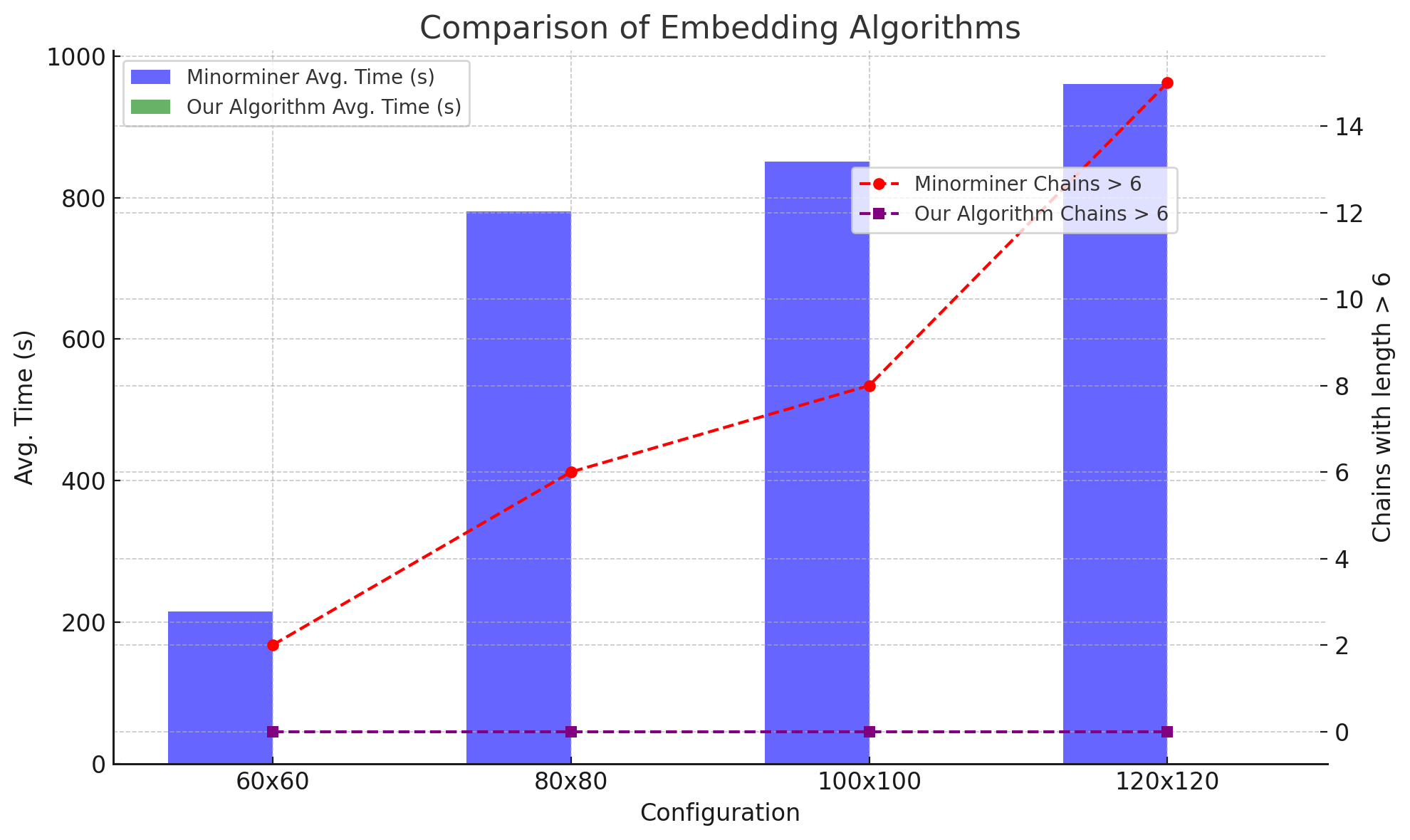}
\label{fig:comparison_embedding}
\end{figure}

\begin{table}[ht!]
\centering
\caption{Comparison of Embedding Algorithms}
\label{tab:embedding_comparison}
\small
\begin{tabular}{|c|p{2.8cm}|c|c|}
\hline
\textbf{Config} & \textbf{Metric} & \multicolumn{2}{c|}{\textbf{Algorithm}} \\ \hline
 &  & Minorminer & Our Alg. \\ \hline
\multirow{3}{*}{60$\times$60}  
    & Avg. Time (s)  & 214.94  & 0.012  \\ \cline{2-4} 
    & Std. Dev (s)  & 108.75  & 0.0012  \\ \cline{2-4} 
    & Avg. Chains $>$ 6  & 2  & 0  \\ \cline{2-4} 
    & Std. Chains $>$ 6  & 1.22  & 0  \\ \hline
\multirow{3}{*}{80$\times$80}  
    & Avg. Time (s)  & 780.65  & 0.012  \\ \cline{2-4} 
    & Std. Dev (s)  & 123.87  & 0.0013  \\ \cline{2-4} 
    & Avg. Chains $>$ 6  & 6  & 0  \\ \cline{2-4} 
    & Std. Chains $>$ 6  & 1.10  & 0  \\ \hline
\multirow{3}{*}{100$\times$100}  
    & Avg. Time (s)  & 850.45  & 0.014  \\ \cline{2-4} 
    & Std. Dev (s)  & 143.54  & 0.0012  \\ \cline{2-4} 
    & Avg. Chains $>$ 6  & 8  & 0  \\ \cline{2-4} 
    & Std. Chains $>$ 6  & 1.17  & 0  \\ \hline
\multirow{3}{*}{120$\times$120}  
    & Avg. Time (s)  & 960.02  & 0.014  \\ \cline{2-4} 
    & Std. Dev (s)  & 170.35  & 0.0011  \\ \cline{2-4} 
    & Avg .Chains $>$ 6  & 15  & 0  \\ \cline{2-4} 
    & Std. Chains $>$ 6  & 1.15  & 0  \\ \hline
\end{tabular}
\end{table}

\section{Conclusions}
In this work, we have introduced an optimized minor-embedding framework for mapping large-scale complete bipartite graphs such as RBMs onto adiabatic quantum processors. Our approach improves upon existing methodologies by leveraging topographical periodicity that efficiently utilises the physical qubit connectivity of state-of-the-art quantum annealers, particularly those based on the Pegasus topology. 

We demonstrated that our embedding technique significantly reduces the length of qubit chains compared to traditional minor-embedding approaches such as Minorminer, leading to enhanced robustness and reduced susceptibility to chain breakage. Our empirical results further confirm that our method enables the embedding of larger RBMs while maintaining computational efficiency, as indicated by the drastic reduction in embedding time and the elimination of long qubit chains.

Moreover, by preserving the bipartite structure of RBMs and ensuring a systematic logical-to-physical mapping, our approach enhances the scalability of quantum machine learning models on quantum annealers. This advancement is particularly relevant for quantum generative modelling, anomaly detection, and hybrid quantum-classical optimization applications.

Future work will explore extending our embedding framework to alternative quantum hardware topologies, incorporating hybrid optimization techniques to refine qubit allocation further, and integrating these embeddings into real-world quantum-enhanced machine learning workflows. Additionally, leveraging advances in error correction and improved qubit coherence times will further enhance the applicability of our embedding framework to large-scale practical problems.

\bibliographystyle{IEEEtran}
\bibliography{references}

\appendix
\section{Supplementary Information}

\subsection{Algorithms}
\begin{breakablealgorithm}
\caption{Embedding a Complete Bipartite Graph}
\label{alg:embedding_graph}
\begin{algorithmic}[1]
\State \textbf{Input:} $n\_visible$, $n\_hidden$, $periodicity\_v$, $periodicity\_h$, $n\_periodicity$
\State \textbf{Output:} Logical-physical mappings for visible and hidden nodes, coupling matrix $J$

\State $H\_V \gets n\_visible / n\_periodicity$ 
\State $H\_H \gets n\_hidden / n\_periodicity$ 
\State Initialize $J\_coupling$, $visible\_nodes$, $hidden\_nodes$, and $J\_connections$ as empty lists

\State $startv \gets periodicity\_v$ 
\For{$z \gets 0$ to $H\_V - 1$}
    \For{$x \gets 0$ to $n\_periodicity - 1$}
        \State $n \gets startv + (n\_periodicity \cdot x) + (z \cdot periodicity\_v)$
        \State Append $n$ to $visible\_nodes$
        \For{$j \gets 0$ to $H\_H - 2$}
            \State Append $(n + j, n + j + 1)$ to $J\_coupling$
        \EndFor
    \EndFor
\EndFor

\State $starto \gets periodicity\_h$ 
\For{$z \gets 0$ to $H\_H - 1$}
    \For{$x \gets 0$ to $n\_periodicity - 1$}
        \State $p \gets starto + (n\_periodicity \cdot x) + (z \cdot periodicity\_h)$
        \State Append $p$ to $hidden\_nodes$
        \For{$j \gets 0$ to $H\_V - 2$}
            \State Append $(p + j, p + j + 1)$ to $J\_coupling$
        \EndFor
    \EndFor
\EndFor

\For{$x \gets 0$ to $H\_H - 1$}
    \For{$y \gets 0$ to $H\_V - 1$}
        \For{$t \gets 0$ to $n\_periodicity - 1$}
            \State $n \gets startv + (n\_periodicity \cdot t) + y \cdot periodicity\_v + x$
            \For{$k \gets 0$ to $n\_periodicity - 1$}
                \State $p \gets starto + (n\_periodicity \cdot k) + x \cdot periodicity\_h + y$
                \State Append $(n, p)$ to $J\_connections$
            \EndFor
        \EndFor
    \EndFor
\EndFor

\State $J \gets 0$
\For{Each $(a, b)$ in $J\_coupling$}
    \State $J[a, b] \gets -1$
\EndFor

\State \Return $J$, $visible\_nodes$, $hidden\_nodes$, $J\_coupling$, $J\_connections$
\end{algorithmic}
\end{breakablealgorithm}

\end{document}